\documentclass[revtex4, astrosymb]{emulateapj}%
\usepackage{graphicx}
\begin{abstract}
An increasing number of non-terminal giant eruptions are being observed by modern supernova and transient surveys.  But very little is known about the origin of these giant eruptions and their progenitors, many of which are presumably very massive, evolved stars.  Motivated by the small number of progenitors positively associated with these giant eruptions, we have begun a survey of the evolved massive star populations in nearby galaxies.  The nearby, nearly face on, giant spiral M101 is an excellent laboratory for studying a large population of very massive stars.  In this paper, we present $BVI$ photometry obtained from archival \textit{HST}/ACS WFC images of M101.  We have produced a catalog of luminous stars with photometric errors $<10\%$ for $V < 24.5$ and $50\%$ completeness down to $V \sim 26.5$ even in regions of high stellar crowding.  Using color and luminosity criteria we have identified candidate luminous OB type stars and blue supergiants, yellow supergiants, and red supergiants for future observation.  We examine their spatial distributions across the face of M101 and find that the ratio of blue to red supergiants decreases by two orders of magnitude over the radial extent of M101 corresponding to 0.5 dex in metallicity.  We discuss the resolved stellar content in the giant star forming complexes NGC 5458, 5453, 5461, 5451, 5462, and 5449 and discuss their color-magnitude diagrams in conjunction with the spatial distribution of the stars to determine their spatio-temporal formation histories.
\end{abstract}

\begin{document}
\title{The Massive Star Population in M101. I. The Identification and Spatial Distribution of the Visually Luminous Stars.}
\author{Skyler Grammer and Roberta M. Humphreys}
\affil{Minnesota Institute for Astrophysics, 116 Church St SE, University of Minnesota, Minneapolis,
MN 55455, USA; grammer@astro.umn.edu, roberta@umn.edu}
\maketitle

\section{Introduction}
Most massive stars will ultimately end their lives as core collapse supernovae however there are several famous examples where massive stars have experienced spectacular non-terminal giant eruptions that rival true supernovae. The modern supernova surveys have produced an increasing number of these non-terminal optical transients with a wide range of properties.  Many of these objects are initially mis-classified as supernovae due to peak luminosities approaching the lower limit for core-collapse supernovae. Their spectral features are also similar to the Type IIn supernovae arising from the strong interaction between the ejecta and the circumstellar medium from prior mass loss episodes \citep{Turatto:1993}. Subsequent observations reveal that these objects are sub-luminous and that their temporal spectral and photometric evolution is not typical of true supernovae. Consequently, these events are sometimes called ``SN impostors'' (\cite{Van-Dyk:2012} and references therein). In some cases the apparent terminal eruption is preceded by smaller eruptions e.g. SN2005gl \citep{Gal-Yam:2007, Gal-Yam:2009}, SN2006jc \citep{Pastorello:2007} and most recently the peculiar SN2009ip \citep{Mauerhan:2013, Pastorello:2013, Fraser:2013, Margutti:2013}.  This of course raises questions about the origin of these giant eruptions and their possible relation to true supernovae.
\begin{figure}
\centering
\includegraphics[width=\columnwidth]{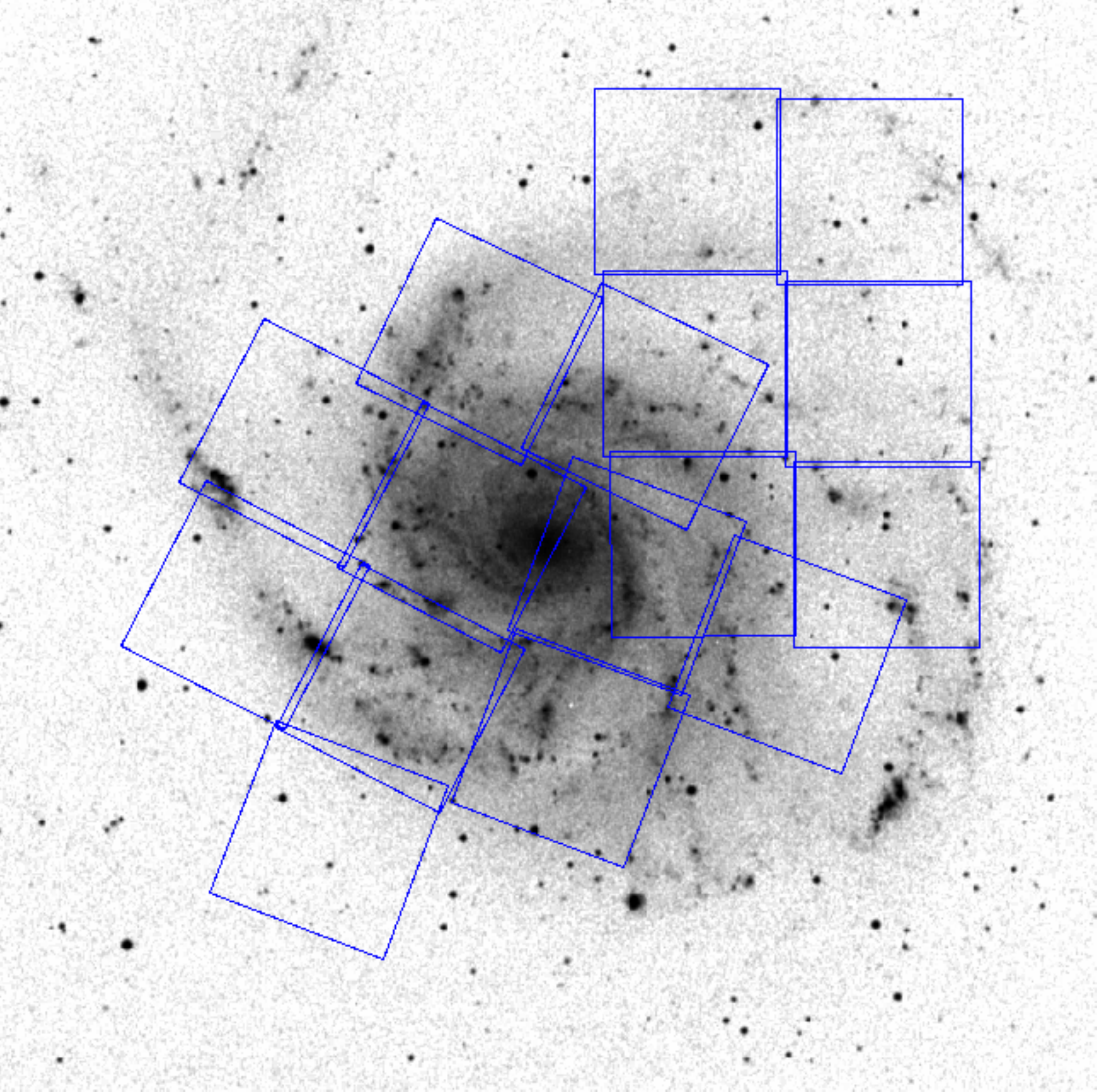}
\caption{POSS-II red image of M101 from the Digitized Sky Survey. The \textit{HST}-ACS fields are shown overlayed onto the image.}
\label{fig:M101view}
\end{figure}
The physical mechanism that triggers these giant eruptions is not known and very little is known about the progenitors. They may come from a range of initial masses and different evolutionary states. An improved census of the likely progenitor classes is now needed; i.e. the evolved most massive stars, the Luminous Blue Variables (LBVs), and the hypergiant stars.  The work presented here is part of a larger survey of the evolved massive star populations in nearby galaxies \citep{Humphreys:2013a}.
\begin{deluxetable*}{c|c|c|c|c|c}[hbt!]
\tabletypesize{\tiny}
\tablecolumns{6}
\tablecaption{HST/ACS Observations \label{tab:HSTobs}}
\tablewidth{0pc}
\tablehead{\colhead{Field Name} & \colhead{$\alpha_{J2000}$} & \colhead{$\delta_{J2000}$} & \colhead{Filters} & \colhead{t$_{exp}$(s)} & \colhead{m$_{50\%}$}} \\
\startdata
9490$\_$01 & 14:03:22.61 & +54:21:27.98 & F435W, F555W, F814W & 2$\times$450, 2$\times$360, 2$\times$360 & 26.8, 26.2, 26.0 \\
9490$\_$02 & 14:03:02.15 & +54:20:36.80 & F435W, F555W, F814W & 2$\times$450, 2$\times$360, 2$\times$360 & 26.9, 26.3, 25.9 \\
9490$\_$03 & 14:03:20.05 & +54:24:51.49 & F435W, F555W, F814W & 2$\times$450, 2$\times$360, 2$\times$360 & 27.0, 26.3, 26.0 \\
9490$\_$a1 & 14:03:30.06 & +54:18:31.08 & F435W, F555W, F814W & 2$\times$450, 2$\times$360, 2$\times$360 & 27.0, 26.4, 26.1 \\
9490$\_$a2 & 14:03:09.05 & +54:17:28.24 & F435W, F555W, F814W & 2$\times$450, 2$\times$360, 2$\times$360 & 27.0, 26.4, 26.1 \\
9490$\_$a3 & 14:02:59.92 & +54:23:43.48 & F435W, F555W, F814W & 2$\times$450, 2$\times$360, 2$\times$360 & 26.9, 26.3, 26.1 \\
9490$\_$b1 & 14:03:49.57 & +54:20:00.69 & F435W, F555W, F814W & 2$\times$450, 2$\times$360, 2$\times$360 & 27.2, 26.6, 26.3 \\
9490$\_$b2 & 14:03:38.87 & +54:15:45.84 & F435W, F555W, F814W & 2$\times$450, 2$\times$360, 2$\times$360 & 27.1, 26.4, 26.1 \\
9490$\_$c1 & 14:03:42.32 & +54:22:59.34 & F435W, F555W, F814W & 2$\times$450, 2$\times$360, 2$\times$360 & 27.0, 26.4, 26.2 \\
9490$\_$c2 & 14:02:41.90 & +54:19:13.46 & F435W, F555W, F814W & 2$\times$450, 2$\times$360, 2$\times$360 & 26.8, 26.3, 26.2 \\
9492$\_$09 & 14:02:53.70 & +54:27:35.70 & F435W, F555W, F814W & 3$\times$360, 3$\times$360, 3$\times$360 & 27.4, 26.8, 26.7 \\
9492$\_$10 & 14:02:52.53 & +54:24:17.80 & F435W, F555W, F814W & 3$\times$360, 3$\times$360, 3$\times$360 & 27.3, 26.6, 26.5 \\
9492$\_$11 & 14:02:51.42 & +54:20:59.90 & F435W, F555W, F814W & 3$\times$360, 3$\times$360, 3$\times$360 & 27.3, 26.6, 26.4 \\
9492$\_$12 & 14:02:30.78 & +54:27:25.88 & F435W, F555W, F814W & 3$\times$360, 3$\times$360, 3$\times$360 & 27.4, 26.8, 26.6 \\
9492$\_$13 & 14:02:29.60 & +54:24:07.97 & F435W, F555W, F814W & 3$\times$360, 3$\times$360, 3$\times$360 & 27.3, 26.6, 26.4 \\
9492$\_$14 & 14:02:28.42 & +54:20:50.06 & F435W, F555W, F814W & 3$\times$360, 3$\times$360, 3$\times$360 & 27.3, 26.8, 26.5
\enddata
\end{deluxetable*}

The nearby, face-on, giant spiral M101 provides an ideal laboratory for the study of luminous massive stars.  Beginning with the early work of \cite{Sandage:1974c} on the brightest blue stars, the vast population of luminous massive stars in M101 has been a subject of interest for the last 40 years \citep{Humphreys:1980, Sandage:1983, Humphreys:1983, Humphreys:1986, Humphreys:1987}. With the launch of \textit{The Hubble Space Telescope} in 1990, the study of the resolved stellar content of M101 became possible and since then many studies have examined and characterized the immense population of Cepheid variables (\cite{Shappee:2011bi} and references therein), stellar clusters \citep{Bresolin:1996, Chen:2005wc, Barmby:2006}, H\,{\small II} regions \citep{Rosa:1994, Pleuss:2000, Garcia-Benito:2011, Sun:2012}, and supernova remnants \citep{Lai:2001, Franchetti:2012}.  The next logical step is the identification of its large population of luminous stars including OB type stars, blue supergiants, yellow supergiants, and red supergiants.  In a similar survey, \cite{Shara:2013} are using narrowband imaging of M101 to identify emission line stars such as the Wolf-Rayet (WR) stars.  

In this first paper we discuss the creation of a catalog of candidate luminous stars.  In the next section, we describe the observations, data reduction, and photometry.  In $\S3$ we discuss the identification of luminous stars of different types.  We explore their spatial distributions and how their relative numbers vary with radius in $\S4$.  In $\S5$, we examine the massive star content within the giant star forming complexes NGC 5458, 5453, 5461, 5451, 5462, and 5449.  We summarize our conclusions in the last section.  Subsequent papers will be dedicated to the discussion of the massive star formation history in M101 (Paper II), the variability and evidence for instabilities in several of our massive stars (Paper III), and the spectroscopy of several of the most luminous and variable stars in M101 (Paper IV). 

\section{Observations and Photometry}
Sixteen fields in M101 were imaged with \textit{The Hubble Space Telescope (HST)} Advanced Camera for Surveys (ACS) Wide Field Camera (Figure~\ref{fig:M101view}) November 13-16 2002 and January 14-23 2003 (proposal IDs 9490 and 9492). In Table~\ref{tab:HSTobs}, we list the field centers, observation dates, filters, exposure times, and 50\% completeness magnitudes from the artificial star tests (discussed below).  Each exposure was bias subtracted and flat-fielded by the STScI On-the-Fly-Reprocessing system.

We used DOLPHOT, a modified version of HSTphot \citep{Dolphin:2000}, with the ACS module to measure the photometry.  DOLPHOT performs the photometry by fitting the ACS point-spread-function (PSF), calculated with TinyTim\footnotemark[1], to all sources in each individual frame.  The PSF magnitudes are aperture corrected using the most isolated stars in each frame and the results from individual exposures are combined.  Count rates are then converted to magnitudes on the VEGAMAG system.  The \textit{HST}/ACS instrumental magnitudes were transformed to the Landolt \textit{BVI} magnitudes using the equations from \cite{Sirianni:2005}.  We corrected the photometry for foreground extinction using E$(B-V) = 0.01$  \citep{Schlegel:1998} and the Galactic extinction curve from \cite{Cardelli:1989}.   
\footnotetext[1]{http://www.stsci.edu/software/tinytim/}

\begin{figure}[hb!]
\centering
\includegraphics[width=\columnwidth, angle=180]{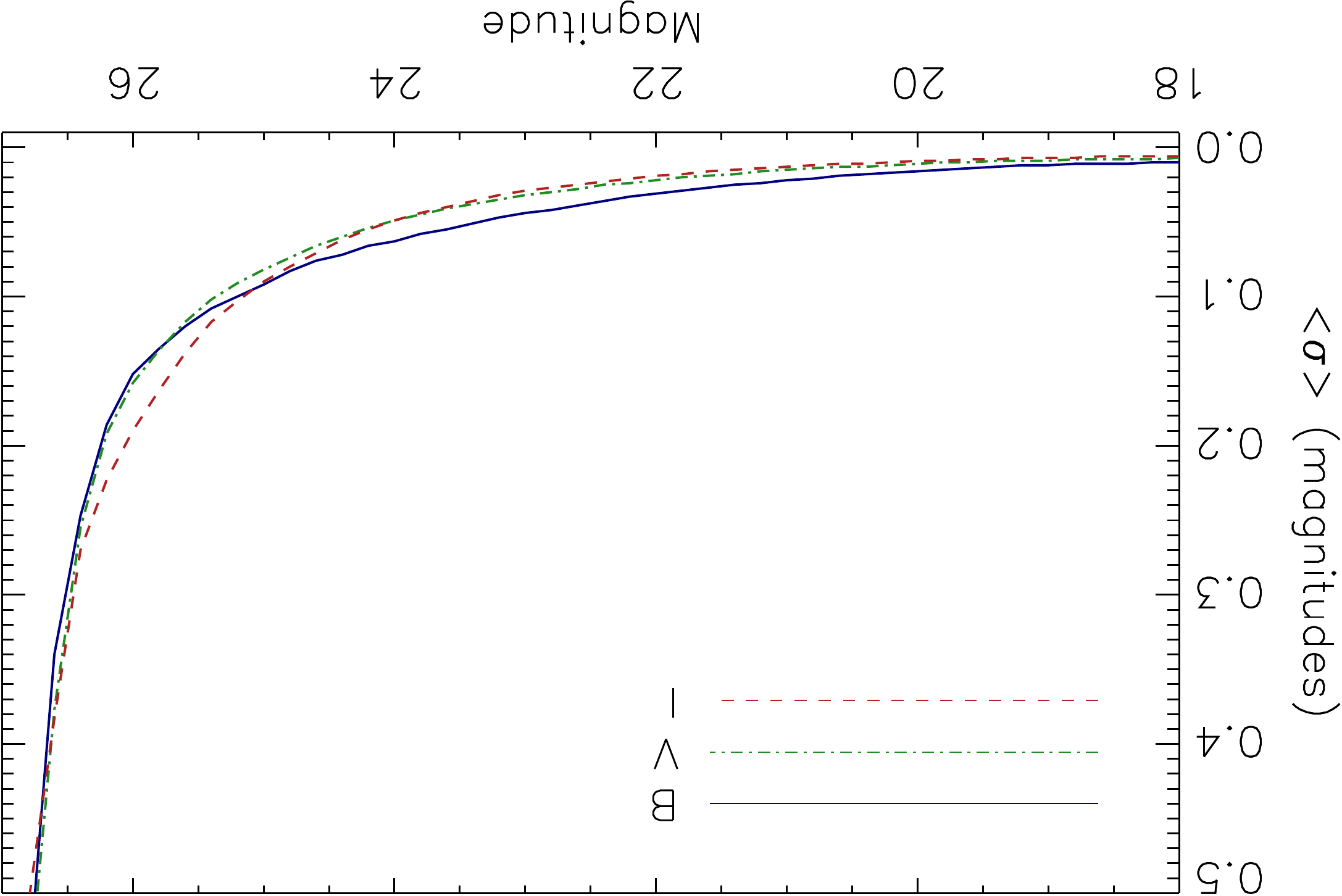}
\caption{Mean photometric errors as a function of magnitude.  Errors are the quadrature sum of the Poisson errors and the photometric error functions derived from the artificial star tests.}
\label{fig:err-functions}
\end{figure}

\begin{deluxetable*}{cccccccccccc}[ht!]
\tabletypesize{\tiny}
\tablecolumns{10}
\tablecaption{M101 Catalog \label{tab:catalog}}
\tablewidth{0pc}
\tablehead{\colhead{ID} & \colhead{Field} & \colhead{$\alpha_{J2000}$} & \colhead{$\delta_{J2000}$} & \colhead{$V$} & \colhead{$\sigma_{V}$} & \colhead{$B-V$} & \colhead{$\sigma_{B-V}$} & \colhead{$V-I$} & \colhead{$\sigma_{V-I}$}}\\
\startdata
J140328.24+541658.12  &  9490$\_$a1  &  14:03:28.24  &  54:16:58.12  &  18.430  &  0.002  &  0.177  &  0.002  &  0.123  &  0.003 \\
J140331.33+542114.40  &  9490$\_$01  &  14:03:31.33  &  54:21:14.40  &  18.490  &  0.002  & -0.060  &  0.002  & -0.067  &  0.003 \\
J140341.49+541904.98  &  9490$\_$b1  &  14:03:41.49  &  54:19:04.98  &  18.552  &  0.002  &  0.188  &  0.003  &  0.371  &  0.003 \\
J140236.66+542145.47  &  9492$\_$14  &  14:02:36.66  &  54:21:45.47  &  18.578  &  0.002  &  0.145  &  0.003  &  0.684  &  0.003 \\
J140248.72+541756.47  &  9490$\_$c2  &  14:02:48.72  &  54:17:56.47  &  18.615  &  0.002  &  0.306  &  0.003  &  0.435  &  0.003 \\
J140334.05+541836.94  &  9490$\_$a1  &  14:03:34.05  &  54:18:36.94  &  18.633  &  0.002  &  0.082  &  0.003  & -0.235  &  0.003 \\
J140312.49+542053.74  &  9490$\_$02  &  14:03:12.49  &  54:20:53.74  &  18.641  &  0.002  &  0.158  &  0.003  &  0.215  &  0.003 \\
J140255.00+542226.94  &  9492$\_$11  &  14:02:55.00  &  54:22:26.94  &  18.670  &  0.002  & -0.015  &  0.002  & -0.056  &  0.003 \\
J140221.40+542541.02  &  9492$\_$13  &  14:02:21.40  &  54:25:41.02  &  18.902  &  0.002  &  1.080  &  0.004  &  1.239  &  0.003 \\
J140301.20+541839.71  &  9490$\_$a2  &  14:03:01.20  &  54:18:39.71  &  18.961  &  0.002  &  0.312  &  0.003  &  0.487  &  0.003
\enddata
\end{deluxetable*}

To transform the raw output of DOLPHOT to a catalog of stellar sources, we used the following parameters: object classification, error flag, signal-to-noise (S/N), sharpness, and crowding.  Sources with non-stellar radial profiles, error flags indicating seriously compromised photometry, and/or S/N $<$ 4 were immediately removed.  We required that the radial profiles be well fit by the ACS PSF using the sharpness parameter and that the photometry not be dramatically affected by neighboring sources using the crowding parameter.  Sharpness is defined to be zero for a perfectly fit PSF, negative for an extended source, and positive for a narrower source (e.g cosmic rays).  Crowding quantifies the degree by which neighboring stars have affected the photometry of an individual star by calculating the difference, in magnitudes, before and after PSF subtraction; large values of the crowding parameter suggest that the photometry may be unreliable.  We experimented with the sharpness and crowding parameters by imposing a range of acceptable values, after which we visually inspected the color-magnitude diagrams (CMD) produced from the accepted and rejected objects.  The final values of sharpness and crowding were adopted when the CMD of rejected objects no longer contained any discernible features, such as a main-sequence or red supergiant branch, and visual inspection of the \textit{HST}/ACS images showed that the accepted objects were not associated with artifacts, diffraction spikes, or background galaxies.  The precise sharpness and crowding criteria that we used to create the final catalog of stellar sources are $|B_{sharp} + V_{sharp} + I_{sharp}| < 0.6$ and $(B_{crowd} + V_{crowd} + I_{crowd}) < 1.5$ magnitudes.

The photometric errors and completeness functions were calculated using artificial stars tests.  We estimated the photometric error functions for each field by injecting 500,000 artificial stars into each frame and then calculated the difference between the input and output magnitudes as a function of input magnitude.  The root-mean-square widths of the error functions were determined at intervals of 0.2 magnitudes and added in quadrature to the Poisson errors generated by DOLPHOT.  Figure~\ref{fig:err-functions} shows the mean error as a function of magnitude for the full catalog.  We find that the photometric errors are less than 10$\%$ for $B < 24.5$, $V < 24.3$, and $I < 24.0$. The completeness of our catalog as a function of magnitude, color, and position was determined by subjecting the artificial star photometry to the same quality criteria as described above.  Color and position averaged completeness functions were generated for each field and the 50$\%$ completeness magnitudes are given in the final column of Table~\ref{tab:HSTobs}.  The catalog is complete at the 50$\%$ level at $B = 27$, $V = 26.5$, and $I = 26.2$ for the fields from proposal ID 9490, and $B = 27.3$, $V = 26.8$, and $I = 26.4$ for fields from proposal ID 9492, which had longer integration times. 

Very massive stars are found in OB associations and clusters where stellar crowding will be problematic, we therefore examined the photometric errors and completeness in regions of high stellar density.  For example, field 9490$\_$01 contains the center of M101 where crowding is high and photometric precision is bound to suffer.  By isolating stars located in the visually most crowded regions of field 9490$\_$01, we find that the photometric errors are better than 10$\%$ down to $B = 24.1$, $V = 23.8$, and $I = 23.6$ and the 50$\%$ completeness magnitudes are $B = 26.5$, $V = 26.2$, and $I = 25.9$.  Therefore even in the most crowded regions of the galaxy, the photometry is still reliable.     

\begin{figure}[hb!]
\centering
\includegraphics[width=\columnwidth]{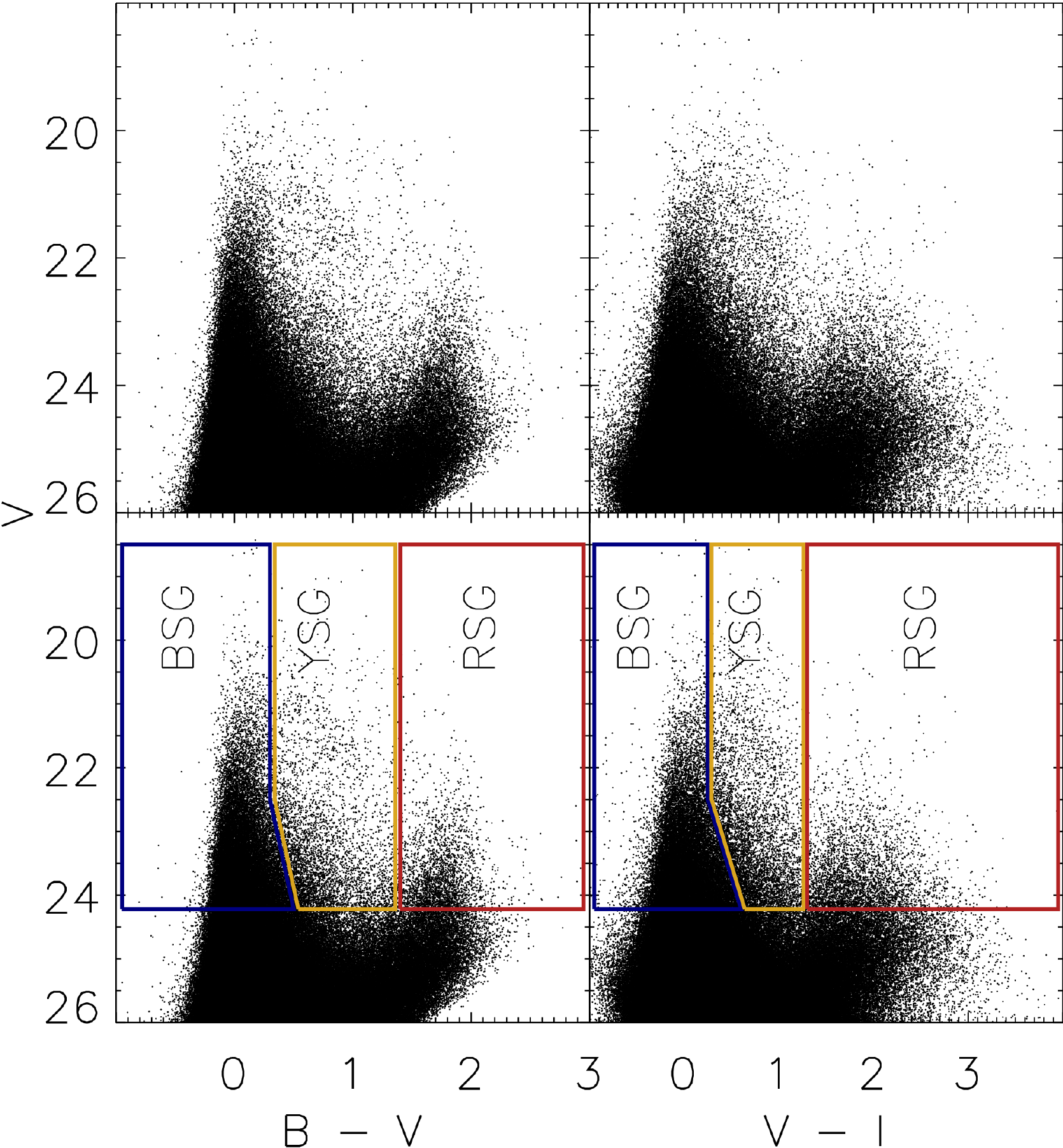}        
\caption{\textit{Top}: Color-magnitude diagrams of M101. \textit{Bottom}: Color-magnitude diagrams showing the selection criteria for luminous massive stars in the BSG, YSG, and RSG subsets (see $\S3$ for details).}
\label{fig:Raw-CMDs}
\end{figure}

WCS positions were converted to the SDSS DR8 FK5 J2000 reference frame by cross-correlating the ACS positions of stars also detected by the SDSS.  An astrometric solution was performed for each frame with errors in the solutions on the order of $0\arcsec.15$.  To remove duplicate entries in the overlapping ACS fields we determined the area of overlap between the two fields and removed stars in the overlapping region from the field with the shorter integration time.  We iteratively applied this process to all field pairs to create the final catalogs for each field.  The individual fields were combined to create the final photometric catalog which contains 539,825 stars in the magnitude range $18.4 < V < 27.6$ and are sorted by increasing $V$ magnitude.

The full version of our catalog is available online but a sample of the catalog is shown in Table~\ref{tab:catalog} which contains the first ten entries.  The following columns are included: identification, ACS field number, right ascension, declination, $V$, $\sigma_{V}$, $(B-V)$, $\sigma_{B-V}$, $(V-I)$, and $\sigma_{V-I}$.  The identifications are based on their $J2000$ coordinates such that J140331.33+542114.40 refers to a star located at $\alpha_{J2000.00}$ = 14:03:31.33 and $\delta_{J2000.00}$ = +54:21:14.40.  The 50$\%$ completeness magnitude of $V\approx26.5$ corresponds to an absolute magnitude of M$_{V} \approx -2.5$ at our adopted distance modulus to M101, $\mu_{0} = 29.05 \pm 0.06 (random) \pm 0.12 (systematic)$, from \cite{Shappee:2011bi}.  In Figure~\ref{fig:Raw-CMDs} we show the CMDs containing all catalog entries.

\begin{figure}[ht!]
\centering
\includegraphics[width=\columnwidth, angle=180]{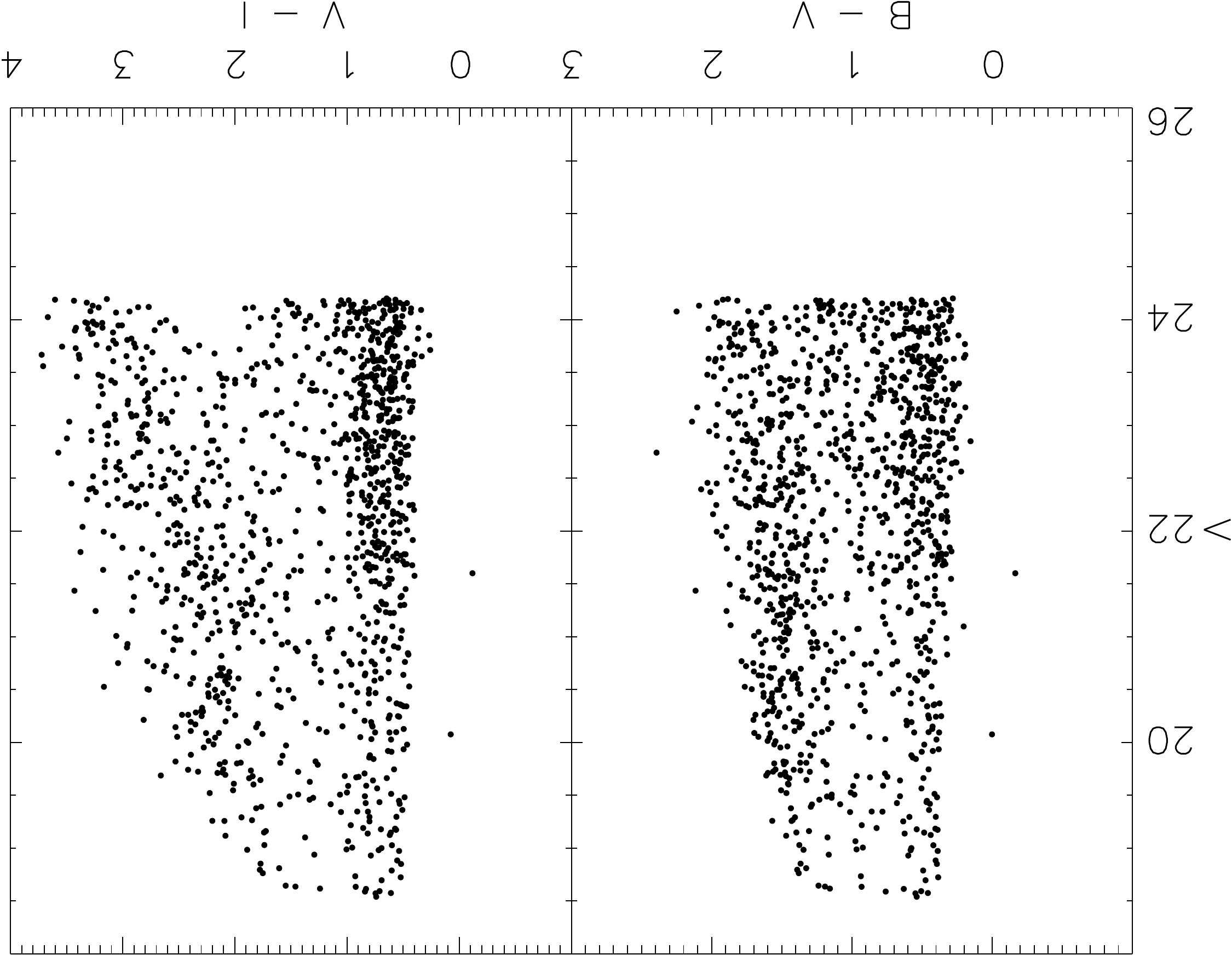}        
\caption{Color-magnitude diagrams of the anticipated Galactic foreground contaminants resulting from the Besan\c{c}on Galactic population synthesis model \citep{Robin:2003} in the direction of M101.}
\label{fig:FG-CMD}
\end{figure}

\section{Identifying the Massive Star Candidates}
To select the different subsets of luminous stars from this very large and comprehensive catalog, we created a subcatalog of stars in the magnitude range $18.5 \leq V \leq 24.2$ where the bright end includes the visually brightest stars in our catalog.  At the distance to M101, the faint limit corresponds to M$_{V} \approx -4.9$, which will include the luminous, hot OB type stars and the most luminous stars at all temperatures.  Fainter than $V = 24.2$ the photometric errors are larger than $10\%$ and are no longer reliable.  This subcatalog includes 35,705 stars.  The CMD in Figure~\ref{fig:Raw-CMDs} shows two well-populated branches which we identify respectively with the blue supergiants and luminous OB type stars, hereafter referred to BSGs, and the red supergiants (RSG) which are located at $(B-V) \geq 1.4$ and $(V-I) \geq 1.3$.  The locus of the BSG group is identified in our CMDs as $(B-V) \leq 0.3$ and $(V-I) \leq 0.25$ for $V < 22.5$.  Fainter than $V = 22.5$, the BSG locus systematically broadens to redder $(B-V)$ and $(V-I)$ colors which may be due to a broadened main-sequence at lower initial masses, interstellar reddening, or to some RSGs evolving back to warmer temperatures.  We therefore adopt somewhat redder $(B-V)$ and $(V-I)$ cutoffs for the magnitude range $22.5 < V \leq 24.2$.  Furthermore we identify the intermediate color, or yellow supergiants (YSGs), as stars with colors in the CMD between the BSG and RSG populations.  In summary our color and magnitude criteria for the BSG, YSG, and RSG populations are given by

\begin{displaymath}
\begin{tabular}{l c r}
{\small BSG} & {\small $(B-V) \leq 0.3$, $(V-I) \leq 0.25$}       & {\small $18.5 \geq V \leq 22.5$} \\
             & {\small $(B-V) \leq 0.5$, $(V-I) \leq 0.6$}        & {\small $22.5 \geq V \leq 24.2$} \\
{\small YSG} & {\small $0.3 < (B-V) < 1.4$}                       & {\small $18.5 \geq V \leq 22.5$} \\
             & {\small $0.25 < (V-I) < 1.3$}                      &                                  \\
             & {\small $0.5 < (B-V) < 1.4 $}                      & {\small $22.5 \geq V \leq 24.2$} \\
             & {\small $0.6 < (V-I) < 1.3 $}                      &                                  \\  
{\small RSG} & {\small $(B-V) \geq 1.4$, $(V-I) \geq 1.3$}        & {\small $18.5 \geq V \leq 22.5$} \\
\end{tabular}
\end{displaymath}

\noindent and are shown superimposed on the CMDs in the bottom half of Figure~\ref{fig:Raw-CMDs}.  

\begin{figure}[htb!]
\centering
\includegraphics[width=\columnwidth, angle=180]{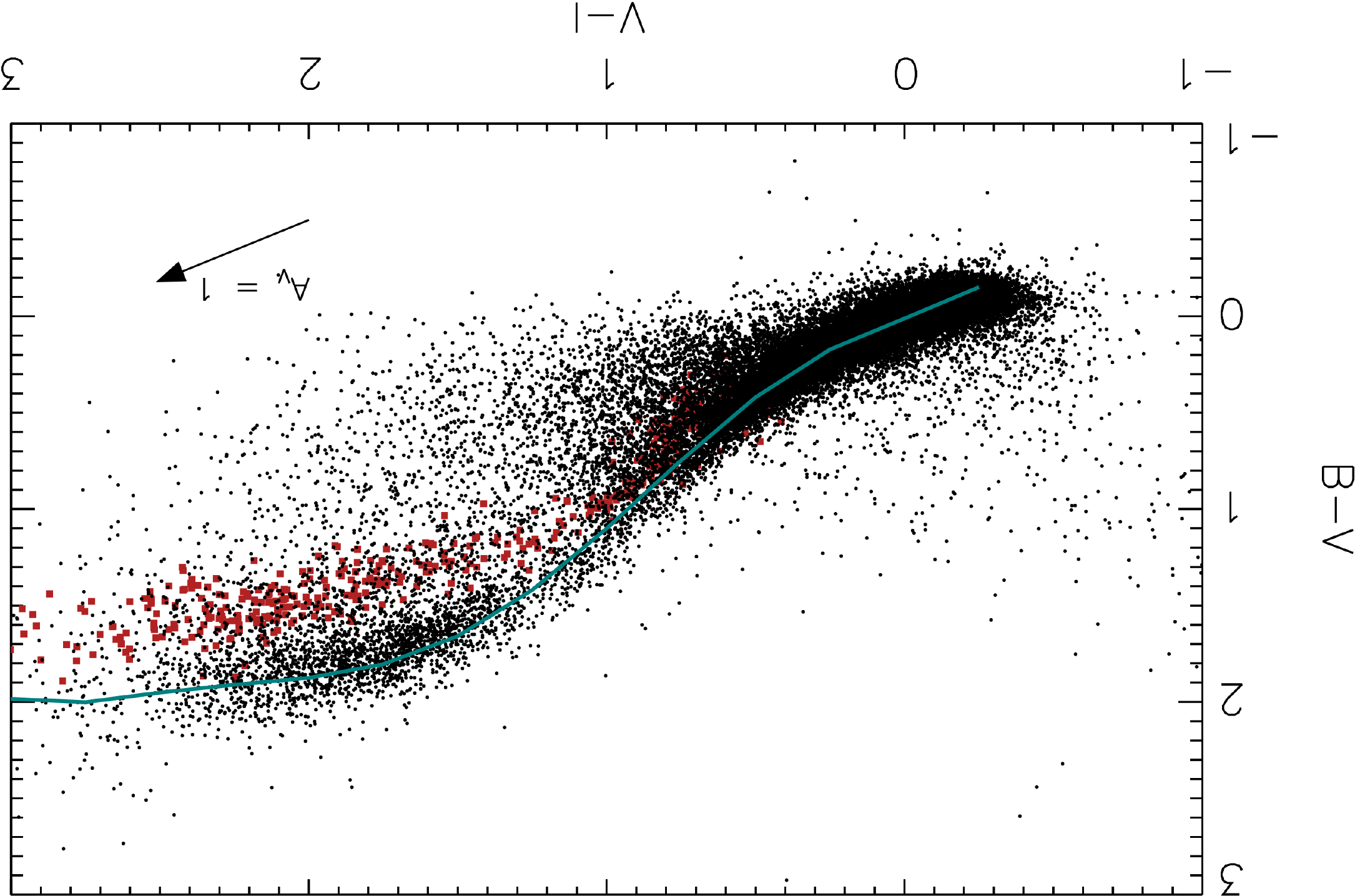}
\caption{Color-color diagram of stars with $V \leq 24.2$ magnitudes. We show the theoretical supergiant sequence from \citep{Bertelli:1994} superimposed as a solid line.  Stars from the Besan\c{c}on Galactic population synthesis model with $V \leq 24.2$ are shown as red squares.  Additionally we show the reddening resulting from A$_{V}$ = 1 assuming a Galactic extinction law \citep{Cardelli:1989}}
\label{fig:SG-CC}
\end{figure}

The use of broad band photometry to identify luminous massive star candidates from our subcatalog increases the likelihood that contaminating sources may be included.  Foreground contamination is commonly estimated by observing a field that is offset from the galaxy, however no such archival field in the same filter set was available for M101.  Consequently we have modeled the Galactic line-of-sight contributions using the Besan\c{c}on Galactic population synthesis model \citep{Robin:2003} over the area covered by the ACS fields.  We used our photometric error functions (Figure~\ref{fig:err-functions}) in the model to obtain the most realistic CMDs of likely foreground stars stars shown in Figure~\ref{fig:FG-CMD}.  As expected, the F through M dwarfs and red giant branch stars make up the largest foreground contribution with colors that are consistent with our defined YSG and RSG groups.  \cite{Massey:1998} showed that RSGs can be photometrically separated from dwarfs in a color-color diagram due to lower surface gravity resulting in redder $(B-V)$ colors for a given $(V-I)$.  We take a similar approach and use a color-color diagram to differentiate probable massive stars from other sources of contamination.  In Figure~\ref{fig:SG-CC} we show the $(B-V)$ versus $(V-I)$ diagram for our subcatalog and model stars with $V \leq 24.2$.  In addition we show the theoretical luminosity class I sequence, hereafter the supergiant sequence, from \cite{Bertelli:1994} as a solid line.  While there is some scatter in Figure~\ref{fig:SG-CC}, stars from the subcatalog form a clear sequence that closely follows the supergiant sequence.  Stars from the Besan\c{c}on model also form a sequence, hereafter called the Besan\c{c}on sequence, that deviates from the supergiant sequence redward of $(V-I) \approx 1$.  Assuming that the Besan\c{c}on model is an accurate representation of the foreground component, we find that the RSGs are separable from foreground stars on a $(B-V)$ versus $(V-I)$ diagram whereas the YSGs are not.  

Asymptotic Giant Branch (AGB) stars are known to overlap in color and luminosity with the RSGs \citep{Brunish:1986}.  However our imposed magnitude limit of $V \leq 24.2$, corresponding to M$_{V} \lesssim -4.9$, is more luminous than the theoretical visual upper luminosity limit for AGB stars at Z = 0.004 \citep{Bertelli:1994}.  At higher metallicities the AGB visual upper luminosity limit is fainter thereby reducing the likelihood that AGB stars populate our RSG sample.

The presence of stars below the Besan\c{c}on sequence in Figure~\ref{fig:SG-CC} and those with very blue $(V-I)$ and red $(B-V)$ colors have photometric properties that cannot be explained by foreground contamination and are likely a mixture of blended objects, unresolved clusters, and reddened stars.  \cite{Barmby:2006} found that clusters would appear to be "slightly resolved" at the distance of M101.  While we have removed sources with radial profiles that deviate appreciably from the ACS PSF using the sharpness parameter, very small clusters may be included in our catalog.  Based on the size distribution and luminosity function of the \cite{Barmby:2006} cluster candidates, we expect around 500 may be included in our subcatalog.  \cite{Barmby:2006} show that the cluster candidates span the range $0 < (V-I) < 1.75$ and $0 < (B-V) < 1$.  Cluster candidates in M51 \citep{Bik:2003} and M81 \citep{Chandar:2001a} are similarly distributed in size, luminosity, and color-color space, thus we find it likely that a large fraction of the stars with anomalous colors may be unresolved clusters.  There are $\sim1500$ stars with anomalous colors, roughly a factor of three greater than than the expected number unresolved clusters.  A fraction of the remaining stars with anomalous colors may be accounted for by blended objects, particularly those with very red $(B-V)$ colors and very blue $(V-I)$ colors and vice versa. 

To account for the remaining stars with anomalous colors, we show a fiducial A$_{V} = 1$ reddening vector in Figure~\ref{fig:SG-CC} which corresponds to E$(B-V) = 0.32$ and E$(V-I) = 0.54$ \citep{Cardelli:1989}.  Stars with the intrinsic colors of $(B-V)_{0} < 0.5$ and $(V-I)_{0} < 0.6$, reddened by $0.5-1$ magnitudes, would migrate to the region of the color-color diagram which contains the highest density of anomalously colored stars.  Consequently it is likely that the remaining stars with anomalous colors are moderately reddened blue stars.  Making the assumption that all the stars that do not follow the supergiant sequence (Figure~\ref{fig:SG-CC}) are a mix of unresolved clusters, blended objects, reddened blue stars, and foreground stars, then we may remove a large fraction of the contaminating sources, at the expense of reducing the completeness of the BSG group, by requiring that candidate luminous massive stars follow the supergiant sequence to within a given tolerance.  The observed width of the subcatalog sequence is $\sim$0.5 magnitudes, thus we further constrain our selection criteria by requiring that colors be consistent with the supergiant sequence to within $\pm$0.25 magnitudes.  

We have taken measures to reduce the amount of foreground contamination to our sample of luminous massive star candidates through the use of a color-color diagram, however as mentioned above, the BSGs and YSGs are not well separated from foreground stars.  Therefore we have characterized the degree by which foreground stars likely contribute to each group by subjecting the Besan\c{c}on model stars to our selection criteria and comparing the relative number of stars in each group.  We have identified 25,603 BSGs, 3,105 YSGs, and 2,294 RSGs and based on the  Besan\c{c}on model, we expect the remaining foreground contribution in the BSGs, YSGs, and RSGs to be $\sim0\%$, $\sim20\%$, and $\sim5\%$ respectively.  Considering the brightest stars in each group ($V < 22$), foreground contamination remains very low in the BSGs ($\ll1\%$) but increases dramatically in the YSGs and RSGs.  Approximately $40\%$ of the bright YSGs are likely foreground stars due to degenerate photometric properties in intermediate color dwarfs and supergiants.  
\begin{figure}[hb!]
\begin{center}
\includegraphics[width=\columnwidth]{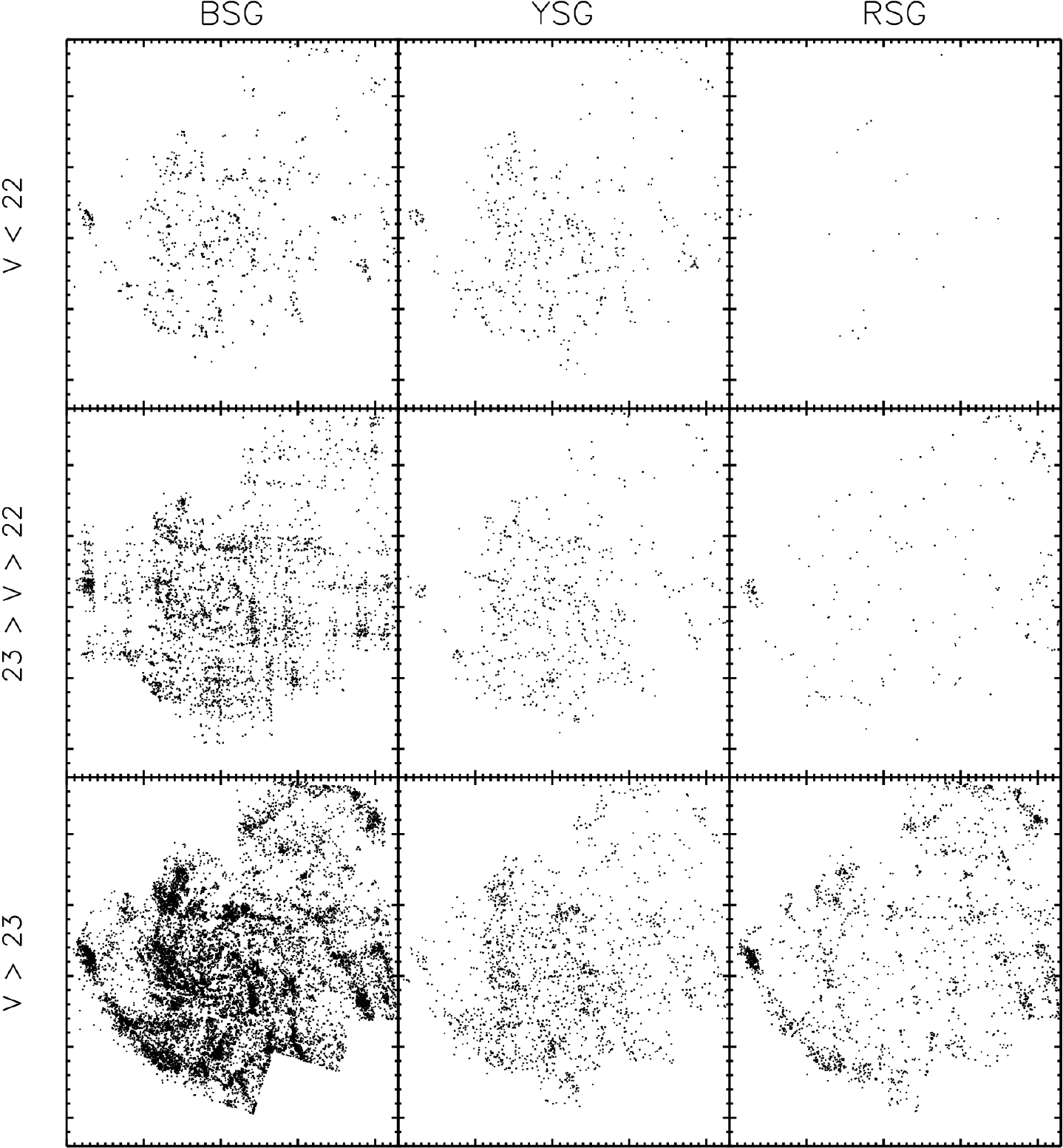}
\caption{Spatial distributions of the luminous massive star candidates with North up and East to the left.  The columns show the candidates separated into the BSG, YSG, and RSG subsets (see $\S3$ for details) and the rows show the dependence on magnitude interval.}
\label{fig:supergiant_map}
\end{center}
\end{figure}

Based on the model, the brightest RSGs are expected to suffer from a large amount of contamination ($50-100\%$) attributed to the overlap between the subcatalog and the Besan\c{c}on stars with $(V-I) > 2$.  However of the 22 brightest RSGs, only 2 have $(V-I) > 2$ therefore the contaminating fraction is likely to be considerably lower for the majority.  We illustrate this point by noting that there are 20 bright RSGs with $(V-I) < 2$ whereas there are 3 stars from the Besan\c{c}on model that have similar photometry.  Consequently the brightest RSGs have a contaminating fraction that is more likely to be $\sim15\%$.  In summary, we find that using our selection criteria $100\%$ of the 25,603 BSGs, $60-80\%$ of the 3,105 YSGs, and $85-95\%$ of the 2,294 RSGs are members of M101.
\begin{figure}[htb!]
\centering
\includegraphics[width=\columnwidth]{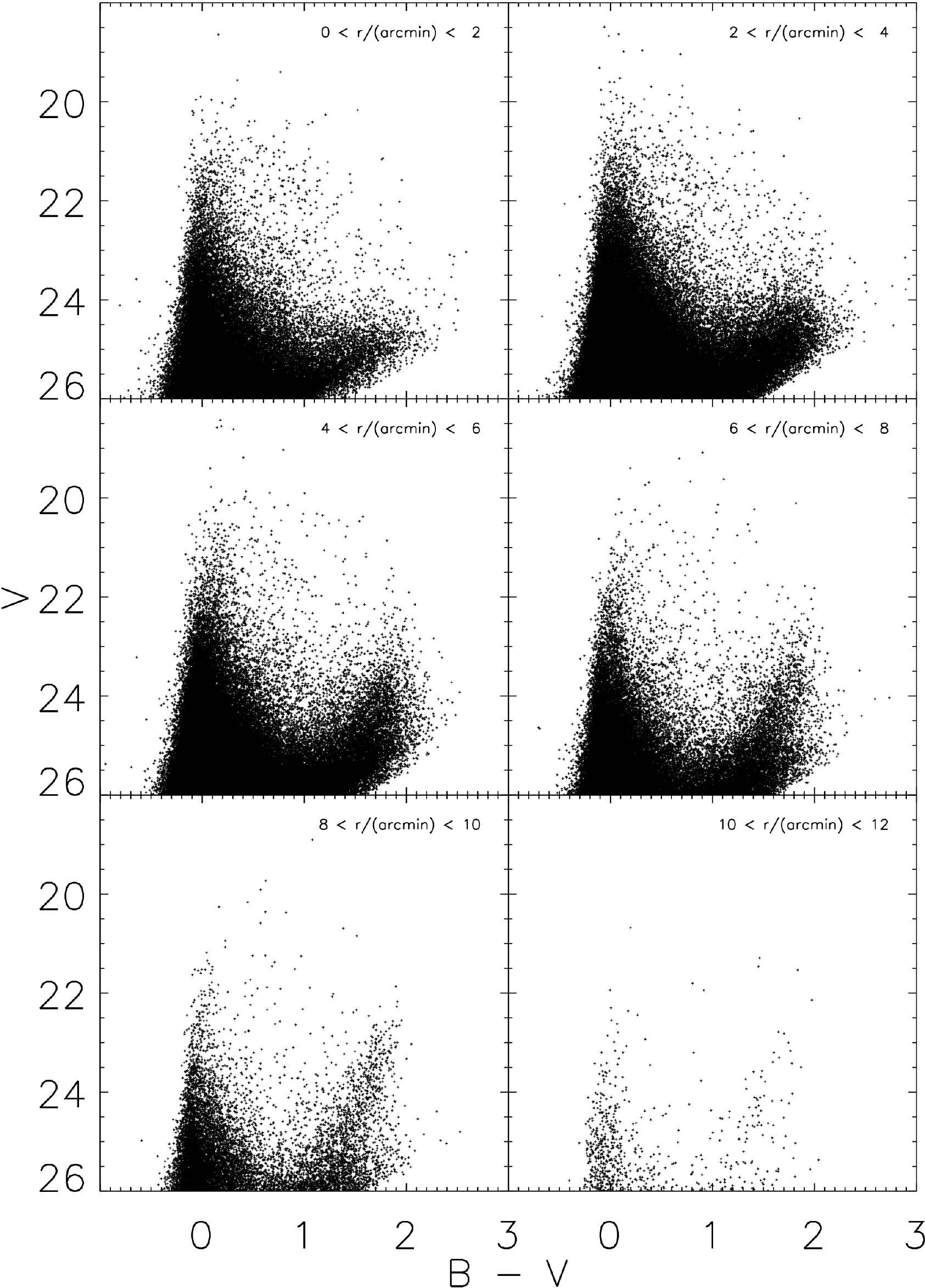}        
\caption{Radial variation of the color-magnitdue diagrams in M101.  The radial distances from the center of M101 were deprojected for each star using an inclination of $18^{\circ}$ and position angle of $39^{\circ}$ \citep{Bosma:1981}.}
\label{fig:radial-CMDs}
\end{figure}

\section{Radial Stellar Population Variations}
A massive, face on, spiral galaxy such as M101 provides an opportunity to examine variations in the stellar content with changes in environment.  M101 is known to have a steep abundance gradient and many regions of intense star formation interspersed throughout that may manifest various effects in the spatial distributions of luminous stars and their CMDs.  First we examine the spatial distributions of the luminous stars in the BSG, YSG, and RSG groups, shown in Figure~\ref{fig:supergiant_map}, which have been separated into three magnitude bins: $V < 22$, $22 \le V < 23$, and $V \geq 23$.  In the top panel of Figure~\ref{fig:supergiant_map}, the brightest BSGs are distributed across the entire face of M101 but show large density enhancements in the spiral arms and massive OB associations.  Though less clumpy and fewer in number, the brightest YSGs appear to have a spatial distribution similar to BSGs.  The brightest RSGs are significantly more scarce and a large fraction may be foreground stars.  In the two fainter bins, the distributions of the luminous massive star candidates are more or less the same except for the RSGs, which become increasingly more prevalent at larger radii.  Adopting an inclination angle of 18$^{\circ}$ and position angle of 39$^{\circ}$ for M101 \citep{Bosma:1981}, we computed the deprojected angular distance, relative to the center of M101, for each star in our catalog and plotted CMDs in six 2$\arcmin$ wide annuli (Figure~\ref{fig:radial-CMDs}).  At a glance, Figure~\ref{fig:radial-CMDs} exhibits the same qualitative behavior seen in the spatial distributions: with increasing radius the RSGs become more prevalent while the number of bright ($V < 22$) BSGs decreases.

The monotonic decrease in the ratio of blue to red supergiants (B/R) with increasing radius and metallcity is a phenomenon that has been well documented by many observers.  \cite{Langer:1995} and \cite{Eggenberger:2002} have reviewed the observational studies of the blue-to-red supergiant ratio.  There are a number of factors that may conribute to the observed ratio.  \cite{Maeder:1980} suggested that the length of time that a star spends as a He burning RSG decreases with increasing metallicity due to enhanced mass loss thereby increasing the time spent as a He burning WR star.  \cite{Massey:1998b} provide support for this scenario by examining the WR and RSG populations in NGC 6822, M33, and M31 showing that the ratio of WR to RSGs monotonically increased with metallicity.  Similarly, studies of the Magellanic clouds, along with OB associations and open clusters in the the Milky Way at various radii, have shown that the B/R ratio displays a marked decrease with metallicity \citep{Meylan:1982, Humphreys:1984, Eggenberger:2002}.

\begin{figure}[hbt!]
\centering
\includegraphics[width=\columnwidth]{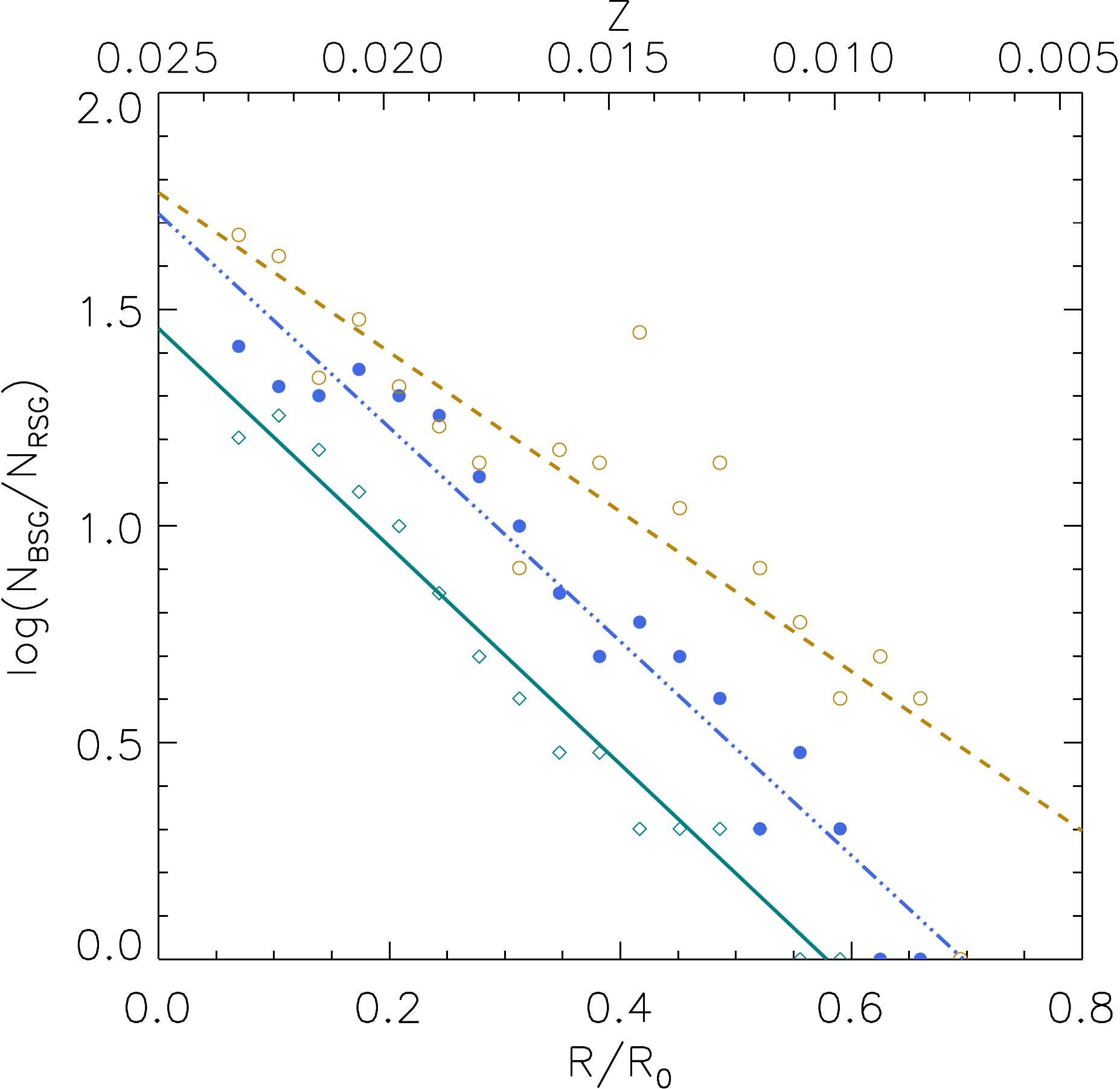}        
\caption{Radial trend of blue-to-red supergiant ratio using the luminous massive star candidates in the BSG and RSG subsets (see $\S3$ for details).  The ratio of BSG to RSG stars have been calculated in three different magnitude intervals in $0\arcsec.5$ bins.  The magnitude intervals are $V < 22$ (gold), $22 < V < 23$ (blue), and $V > 23$ (green), and the best fit lines are shown for each.  Radii have been converted to metallicity using the oxygen abundance gradient from \cite{Kennicutt:2003} assuming a solar oxygen abundance of $12+log(O/H)_{\odot} = 8.69$ \citep{Allende-Prieto:2001}.}
\label{fig:BR_ratio}
\end{figure}

We have quantified the apparent trends in the CMDs by calculating the number ratio of candidate BSGs to RSGs as a function of radius and metallicity for three magnitude intervals: (i) $V < 22$, (ii) $22 \leq V < 23$, and (iii) $V \geq 23$ in bins of 0$\arcmin$.5.  The radii were converted to metallicity using the oxygen abundance gradient from \cite{Kennicutt:2003} and adopting a solar oxygen abundance of 12 + log(O/H) = 8.65 \citep{Allende-Prieto:2001}.  The resulting B/R ratio versus radius and metallicity trends are shown in Figure~\ref{fig:BR_ratio}.  We find that while the relative offsets between each magnitude interval differ, the slopes are comparable and agree to within $<$10$\%$ when comparing the two fainter magnitude intervals (ii) and (iii).  However we note that the slope in the brightest magnitude interval (i) differs by $\sim$25$\%$ when compared to the two fainter ones.  RSGs brighter than $V = 22$ are quite scarce, as seen in Figure~\ref{fig:radial-CMDs}, and our analysis of the foreground contribution indicates that at least $15\%$ may be foreground stars, therefore the B/R ratio may not be trustworthy in this magnitude interval.  Consequently we have taken the average of the slopes from magnitude intervals (ii) and (iii) and find a B/R ratio gradient of $\Delta$log(B/R)/$\Delta$Z = 95$\pm$20 roughly consistent with the $\sim$65 found by \cite{Humphreys:1984}, $\sim$45 by \cite{Meylan:1982}, and $\sim$80 by \cite{Eggenberger:2002}.  We are hesitant to interpret the observed radial decline in the B/R ratio as being solely due to the decreasing metal abundance of M101 given that a spatially varying mean stellar age could also manifest a B/R ratio radial dependence.  In order to disentangle the effects of age and metallicity on the B/R ratio, knowledge of the radial star formation history is required, thus we defer to a future paper a more thorough analysis.   

\section{The Massive Star Content of Giant Star Forming Complexes}
The large star forming complexes, or giant H\,{\small II} regions (GHR), of a galaxy are small-scale examples of extreme star formation that are similar to starburst galaxies.  GHRs contain many young spatially coincident star clusters but the formation history could be coeval or separated by several million years.  The CMD is a powerful tool used to gain insights into the stellar populations present as well as extinction properties.  Studying the visual morphology and spatial locations of stars, in conjunction with the CMDs, provides clues into the formation of the GHRs. In this context we present and discuss the resolved stellar content, which has not been previously studied in detail, within the GHRs NGC 5458, NGC 5453, NGC 5461, NGC 5451, NGC 5462, and NGC 5449.  

We have selected stars from inside a 30$\arcsec$ box centered on each GHR and have created CMDs shown in Figure~\ref{fig:GHR-CMDS}.  Stellar isochrones between the ages $5-100$ Myrs and masses $1-100$ M$_{\odot}$ \citep{Bertelli:1994} and have been included to aid in the interpretation of the stellar content and star formation history.  The GHRs in the inner regions are expected to contain a sizable contribution from the disk which decreases in density exponentially with increasing radius.  We have gauged the degree of disk contribution to the GHR CMDs by extracting stars in 30$\arcsec$ boxes from regions in M101 that do not show any obvious signs of star formation.  While the density of stars decreases with radius, the overall morphology of the CMDs remains more or less the same.  Figure~\ref{fig:Disk-CMD} shows a representative disk CMD at a radius of 3$\arcmin$.  Comparing the CMDs of the GHRs to disk stars, we see that our analysis of the the massive star populations should not be affected by contamination since the density of bright main-sequence/blue supergiants stars is low and the number of stars with $(B-V) > 1.3$ and $V < 25$ is effectively zero. At larger radii, the exponential nature of the disk will further reduce the contamination.   

In addition to CMDs, we have analyzed the spatial distributions of BSGs and RSGs present in each GHR visually and quantitatively, through the use of a two dimensional Kolmogorov-Smirnov (2-D KS) test, to look for spatial variability in the mean stellar age.  We have separated the stars into three groups: $B1, B2$, and $R$.  We have chosen $B1$ and $B2$ to correspond to BSGs with $V < 22$ and $22 < V < 23.5$, respectively.  BSGs in $B1$ only exist on the optical CMD between the ages of 5 and 10 Myrs while those in $B2$ may exist anywhere between 0 and 30 Myrs.  RSGs with $22 < V < 24$ are assigned to $R$ and may be present on the CMD between the ages of 10 and 20 Myrs.  By comparing the spatial distributions of all three groups, we are able to determine the spatio-temporal formation history for each GHR.  We show the spatial distributions of each group in Figure~\ref{fig:GHR-spatial} for visual inspection and make use of the 2-D KS test for a quantitative comparison.  In the 2-D KS test, the test statistic, $D_{KS}$, is defined as the maximum cumulative difference between the empirical $F(\alpha,\delta)$ distribution functions.  The probability of drawing a smaller $D_{KS}$, assuming the two distribution functions are identical (ie. the null hypothesis is true), is given by $100\% \times p$.   

As with any statistical test, it is important to know how various parameters affect the results, thus we have performed Monte Carlo simulations to test the sensitivity of the 2-D KS test to sample size and distribution function parameters such as spatial extent and offset.  Using the same 30$\arcsec$ FOV, we simulated Gaussian spatial distributions and varied the $FWHM$ by $6\arcsec.5$ - 20$\arcsec$, and offset the two distributions by 0$\arcsec$-5$\arcsec$.  We tested several different scenarios where the distribution functions were identical and where they differed by either FHWM, offset, or both.  For each scenario, we performed $10^{4}$ iterations and performed the 2-D  KS test each time.  We found that the 2-D KS test was extremely sensitive to spatial offsets and rejected the null hypothesis at high significance ($p < 0.005$) when identical distributions were offset by $\geq1\arcsec$.  When the distributions were spatially coincident, but differed in $FWHM$, the null hypothesis was reliably rejected to high significance ($p \ll 0.001$ and no greater than $p = 0.05$) when $|1 - FWHM1/FWHM2| \geq 0.1$.  For cases when $|1 - FWHM1/FWHM2| <  0.1$, the 2-D KS test was incapable of telling the distributions apart at high significance and the $p$ values ranged between 0.09 and 0.15.  Furthermore when we forced the distributions to be equal but varied the $FWHM$, we found that $p$ values were always $\geq 0.2$.  Based on our Monte Carlo simulations we conclude that it is safe to assume that two distributions are significantly different when $p \lesssim 0.08$ and identical when $p > 0.2$.  For intermediate values, $0.08 < p < 0.15$, we take a conservative approach and assume that the distributions are similar enough to be considered identical.  When varying the sample sizes, we found that our conclusions held as long as the sample size of each distribution was N $\geq$ 10.  When a distribution had N $<$ 10, the 2-D KS test was no longer capable of separating distributions that were slightly or even moderately different.  

\subsection{NGC 5458}
\begin{figure}
\centering
\includegraphics[width=\columnwidth]{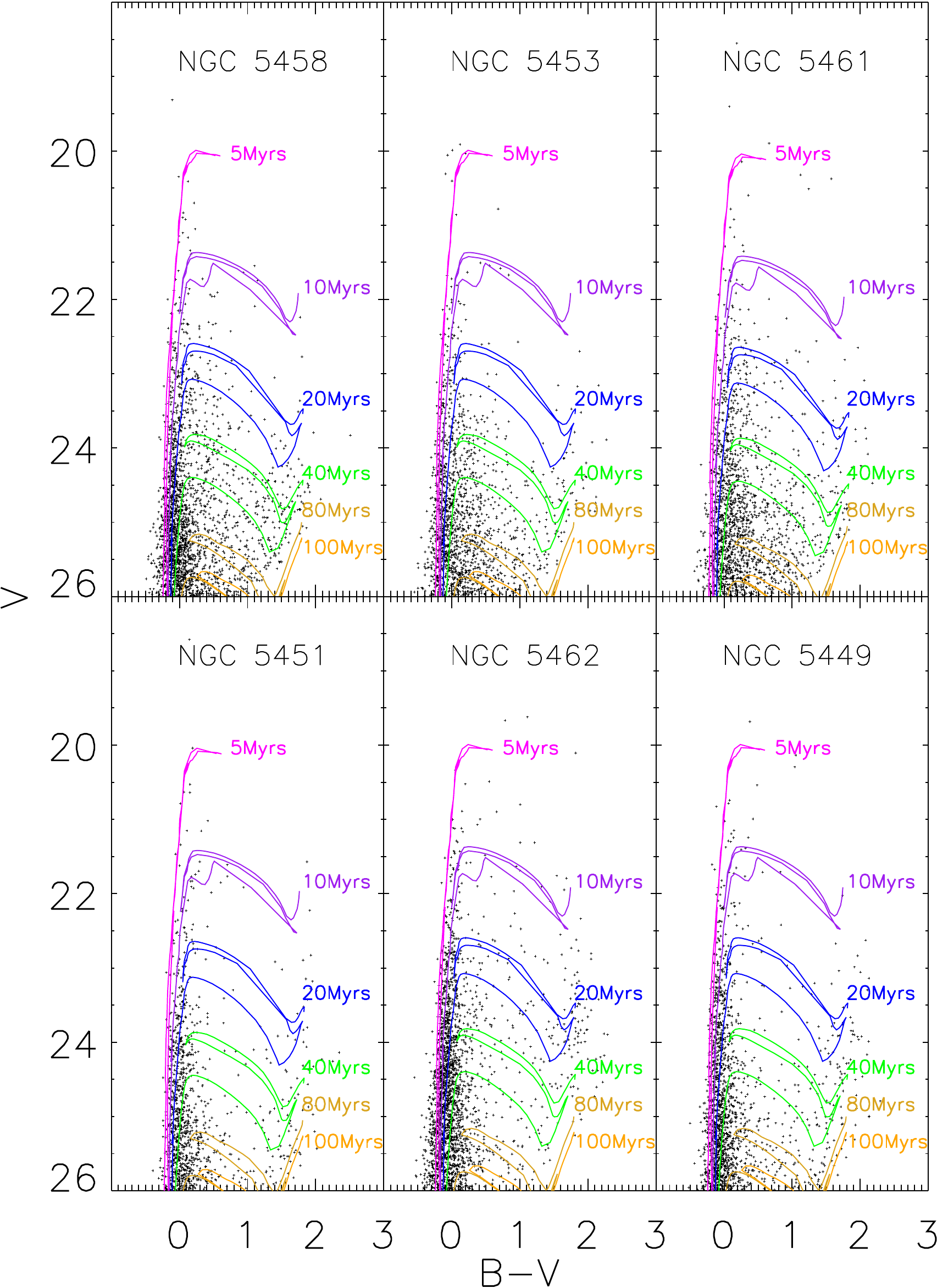} 
\caption{Color-magnitude diagrams for the giant H\,{\small II} regions with NGC designations.  Regions are sorted by increasing distance from the center and have radial angular distances of $3\arcmin.1$, $3\arcmin.5$, $4\arcmin.5$, $5\arcmin.3$, $6\arcmin.0$, and $6\arcmin.6$, respectively.}
\label{fig:GHR-CMDS}
\end{figure}

\begin{figure}
\centering
\includegraphics[width=0.75\columnwidth]{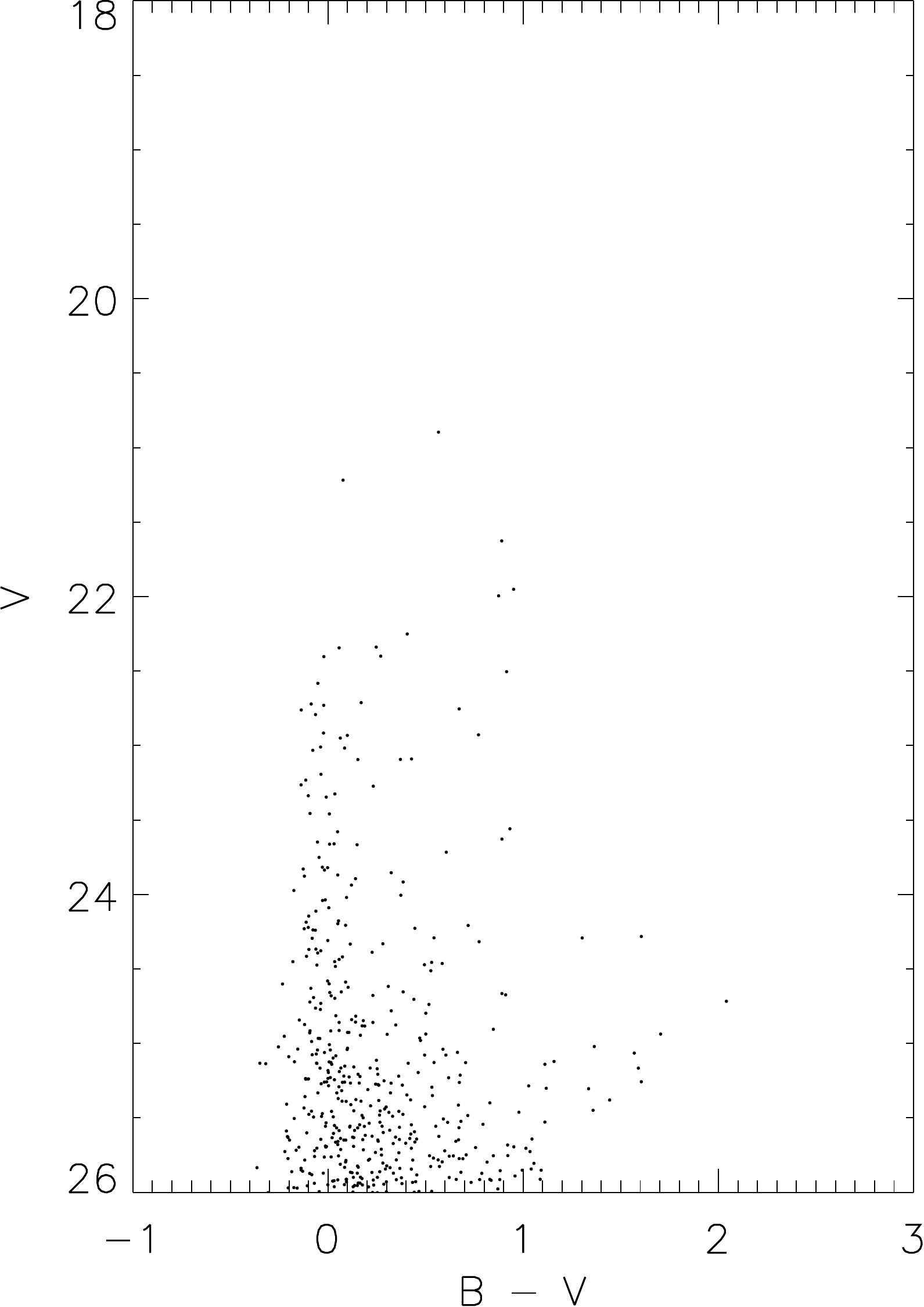}        
\caption{Representative color-magnitude diagram of the disk component at a radius of $3\arcmin$.  At larger radii, the color-magnitude diagrams look similar however there are fewer stars due to the stellar exponential density profile.}
\label{fig:Disk-CMD}
\end{figure}

The CMD of NGC 5458 in Figure~\ref{fig:GHR-CMDS} shows a well defined main-sequence and BSGs extending up to $V \approx 20$ indicative of star formation as recent as 5 Myrs ago.  Integrated spectroscopic observations in the FUV indicate that NGC 5458 is dominated by stellar populations 5.5 to 6.0 Myrs old consisting of 150 O type stars and 18 WR stars \citep{Pellerin:2006wn}.  In addition to a young population of stars, we see that there are red stars ranging between $23.5 < V < 25$ thus NGC 5458 also contains a more evolved component between 20 and 40 Myrs old.  Fainter than $V = 25$, the CMD is dominated by the disk of M101 (Figure~\ref{fig:Disk-CMD}).  The spatial distributions of stars in $B1$, $B2$, and $R$ do not show any obvious differences by eye (Figure~\ref{fig:GHR-spatial}) which is confirmed by the 2-D KS test.  NGC 5458 thus appears to be a single complex of star formation that has been producing stars continuously over the last $\sim$40 Myrs.

\subsection{NGC 5453}
The CMD of NGC 5453 is similar to NGC 5458 although it lacks the population of stars with ages $\sim$5 Myrs.  Eight candidate clusters are located within 30$\arcsec$ of the center of NGC 5453 \citep{Barmby:2006} and account for the brightest blue objects with $V < 20.5$ in Figure~\ref{fig:GHR-CMDS}.  The diffuse X-ray luminosity generated by powerful stellar winds and supernovae is time-dependent and increases by almost two orders of magnitude for ages between 3 and 10 Myrs then disappears for ages greater than 40 Myrs \citep{Oskinova:2005}.  For a single 10$^{6}$ M$_{\odot}$ cluster at 10 Myrs, the X-ray luminosity is log(L$_{X}$/L$_{\odot}$) $\sim$ 3 \citep{Oskinova:2005} which, at the distance of M101, corresponds to a flux that is below the detection threshold of the \textit{CHANDRA} image of M101  \citep{Kuntz:2003} in the hard bandpass and only a few times brighter than the detection threshold of the soft bandpass.  NGC 5453 was undetected by \textit{CHANDRA} \citep{Kuntz:2003} and is consistent with models of cluster X-ray luminosity if the dominant population is $\geq$10 Myrs and the brightest objects are $10-20$ Myr clusters with masses $\lesssim 10^{4}$ M$_{\odot}$.  The spatial distributions of $B1, B2$, and $R$ stars do not differ at high significance leading us to suspect that the star formation in NGC 5453 has been uninterrupted and spatially invariant, similar to NGC 5458.

\subsection{NGC 5461}
NGC 5461 is one of the brightest GHRs within 10 Mpc \citep{Kennicutt:1984} with a surface brightness commensurate to the most active starburst regions and an H$\alpha$ flux on par with the ionizing flux of super star clusters \citep{Kennicutt:1988, Luridiana:2001}.  At a glance the CMD tells us that there are at least two dominant populations, one that is $\leq$5 Myrs old and another that is $>$10 Myrs.  Spectral synthesis of the optical to NUV wavelengths indicate that the dominant stellar population is very young, (3 to 4.5 Myrs) and from the ratio of H$\alpha$ to H$\beta$, extinction is on the order of E($B-V$) = 0.23 \citep{Rosa:1994}.  The FUV spectrum of NGC 5461 is best fit by a stellar population consisting of 175 O stars and 18 WR stars between the ages of 3.3 and 4 Myrs, in agreement with optical data, although the FUV spectrum is best fit by zero reddening \citep{Pellerin:2006wn}.  Integrated FUV observations of a region with variable extinction are likely to under-represent the most reddened stars due to the steep extinction curve from optical to UV \citep{Cardelli:1989} hence the discrepancy in the measured extinction.  The visual image of NGC 5461 shows a filamentary core indicating that there are regions that may be highly reddened.  The width of the main-sequence, at magnitudes where uncertainties are small, supports the proposition of differential reddening on the order of $\Delta$A$_{V} \approx$ 1 magnitude.  

Examining the spatial distributions of stars, we find that $B1$ and $R$ stars are primarily located to the Southwest of the extremely bright H$\alpha$ source H1105 \citep{Hodge:1990} which contains six massive clusters with ages $<$5 Myrs \citep{Chen:2005wc}.  Comparing the $B1$ stars $R$ stars confirms that they arise from the same spatial distribution ($p = 0.3$) whereas the comparison between $B2$ and $R$ exhibits differences that are statistically significant ($p = 0.08$).  Comparing $B1$ to $B2$, the 2-D KS test yields $p = 0.09$ which we are not comfortable interpreting as definitively different given the results of our Monte Carlo tests.  However when we removed the Easternmost $B1$ star, which is much further away from its nearest neighbors than any other $B1$ star, the significance increased to $p = 0.03$ which is well within the our limits for a significant difference.  Combining the results of the 2-D KS test with the CMD and previous studies of NGC 5461, we come to the conclusion that star formation occurred first in the Southwest, $5-20$ Myrs ago, and has propagated to the Northeast to create the youngest stars and clusters which make up H1105. 

\subsection{NGC 5451}
\begin{figure*}
\centering
\includegraphics[width=\textwidth,angle=180]{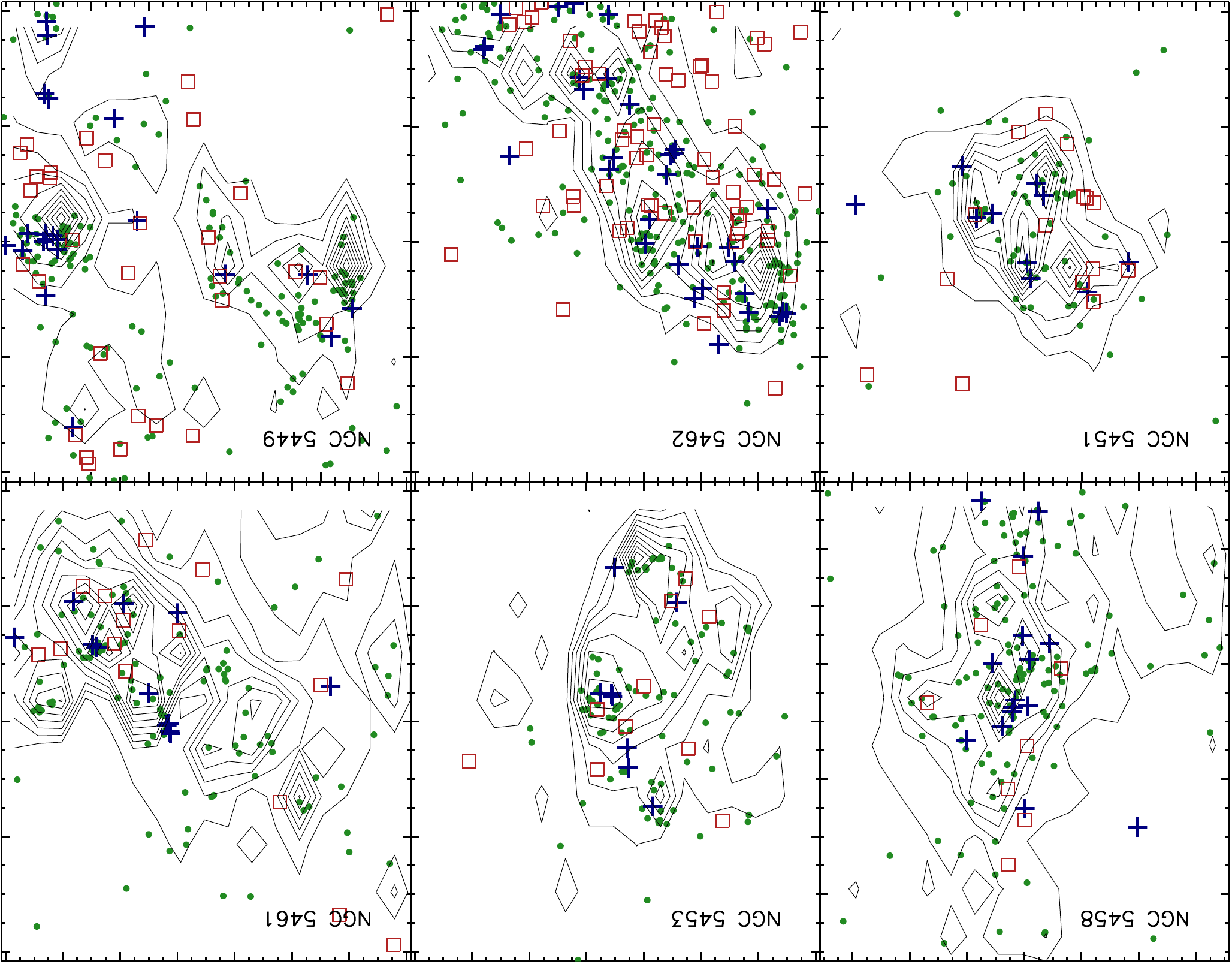}     
\caption{Spatial distribution of stars in the giant H\,{\small II} regions.  The subgroups $R, B1$, and $B2$, which are defined in the text, are shown as red squares, navy crosses, and green points, respectively.  Stellar density contours for stars with $V < 25$ are shown for reference.}
\label{fig:GHR-spatial}
\end{figure*}

The CMD of NGC 5451 (Figure~\ref{fig:GHR-CMDS}) exhibits a clear RSG sequence which extends from $V = 25$ to $V = 21.5$ showing that star formation has been ongoing between 15 and 40 Myrs ago.  The existence of main-sequence stars or BSGs as bright as $V = 20$ along with the presence of several H\,{\small II} regions \citep{Hodge:1990} indicates that star formation is ongoing.  Comparing the spatial locations of blue to red stars in Figure~\ref{fig:GHR-spatial}, we find that the overall distributions do not differ considerably thus star formation in NGC 5451 has been continuous and monolithic within the last 40 Myrs.
\subsection{NGC 5462}
NGC 5462 is one of the largest GHRs in M101 and has been shown to contain as many as 25 young massive clusters \citep{Chen:2005wc}.  Inspection of the CMD (Figure~\ref{fig:GHR-CMDS}) reveals an interesting mix of stellar populations with the youngest stars consistent with ages of $\sim$10 Myrs.  RSGs span roughly three magnitudes starting at $V = 24$ indicating star formation additionally occurred between 15 and 40 Myrs ago.  While constant, the amplitude of the star formation rate may have been extremely high during the period of $15-30$ Myrs ago given that RSGs are significantly overabundant in that age range.  The maximum mass of a cluster and star formation rate are tightly correlated parameters \citep{Larsen:2002}.  \cite{Chen:2005wc} find that the two most massive clusters in NGC 5462 have ages $>$10 Myrs which may support the presence of an epoch of enhanced star formation rate occurring $\geq$10 Myrs ago.  

$B1$ and $B2$ stars are tightly distributed along the major axis of NGC 5462 while the $R$ stars are spread throughout showing the highest density Southeast of the major axis.  The 2-D KS test substantiates these apparent trends to high significance ($p \leq 0.02$) indicating that the $B1$ and $B2$ stars differ in spatial distribution from $R$ stars while $B1$ and $B2$ stars are identically distributed ($p = 0.8$).  NGC 5462 is the brightest GHR in X-rays with a strong diffuse component, due to heating from the stellar winds of extremely massive stars, Southeast and spatially offset from the H\,{\small II} emission in the Northwest \citep{Williams:1995, Wang:1999wl, Kuntz:2003, Sun:2012}.  Visually there exists a clear separation between the stellar concentrations and the nebular emission which in conjunction with the analysis of the spatial distributions of $B1$, $B2$, and $R$ stars, suggests that star formation began in the Southeast between 20 and 30 Myrs ago and has propagated in the Northwest direction subsequently producing the $<$10 Myr population.

\subsection{NGC 5449}
The CMD (Figure~\ref{fig:GHR-CMDS}) of NGC 5449 shows a spread of $\sim$3 magnitudes in the RSG luminosities and the presence of blue stars extending up to $V = 20$.  Star formation over the last $\sim$40 Myrs therefore appears to have been continuous and is ongoing given that there are many H\,{\small II} regions within \citep{Hodge:1990}.  Figure~\ref{fig:GHR-spatial} shows that the spatial distribution of the stars is bimodal and the division of $B1$, $B2$, and $R$ stars is not uniform.  By eye, there appears to be more $B1$ and $R$ stars in the West and when comparing $B1$ stars to $B2$ and $R$ stars, we find that to very high significance ($p \ll 0.01$) the $B1$ stars have a spatial distribution that is significantly different from $B2$ and $R$ stars.  Comparing the $B2$ and $R$ stars, we find that they are statistically identical but not to high significance ($p = 0.15$), therefore the only statistically significant difference between the two concentrations of stars is the overabundance of $B1$ stars in the West.  Furthermore the location of the most luminous H\,{\small II} region within the NGC 5449 star forming complex is to the East \citep{Hodge:1990} and given that the visually brightest blue stars are not the most massive, due to large bolometric corrections associated with high effective temperatures, we suggest that the Western concentration is likely to be more evolved than the Eastern concentration.

\section{Summary and Future Work}
Using archival \textit{BVI} \textit{HST}/ACS images we have created a catalog of luminous stars covering most of the optical disk in M101.  Unprecedented spatial resolution paired with excellent characterization of the \textit{HST}/ACS point spread function and detectors have allowed for the minimization in the numbers of non-stellar sources, such as background galaxies and unresolved clusters, and the ability to maintain high photometric precision and depth even in regions of M101 where stellar crowding is extreme.  Using the catalog we have studied the massive star content of M101 and our main results are the following:

\begin{enumerate}
\item We have identified luminous massive star candidates, for future study, using color-magnitude and color-color diagrams.   We separated the luminous massive star candidates into three subsets: luminous OB type stars and blue supergiants, yellow supergiants, and red supergiants.  We modeled the foreground contamination in the direction of M101 using the Besan\c{c}on Galactic population synthesis model \citep{Robin:2003} and conclude that using our selection criteria, $\sim100\%$ of the 25,603 luminous OB type stars and blue supergiants, $60-80\%$ of the 3,105 yellow supergiants, and $85-95\%$ of the 2,294 red supergiants are members of M101.

\item Examining the spatial distributions of our candidates, we find that the blue and yellow supergiants share a common distribution however the red supergiants are more common in the outer parts of the galaxy, supported by color-magnitude diagrams at various radii.  The blue to red supergiant ratio decreases smoothly with radius and after converting radius to metallicity, we observe a decrease of roughly two orders of magnitude over 0.5 dex in metallicity.  However as mentioned in the text, the dependence on metallicity remains uncertain since a spatially variable star formation history could result in a blue to red supergiant ratio that decreases radially.

\item We discuss the resolved stellar content of giant H\,{\small II} regions NGC 5458, 5453, 5461, 5451, 5462, and 5449.  Through the use of a 2-D Kolmogorov-Smirnov test, we quantitatively compare the spatial distributions of blue and red supergiants in different magnitude intervals in order to determine the spatio-temporal formation history for each region.  We find that in all the star forming regions, the color-magnitude diagrams display evidence for continuous star formation over the last 40 Myrs.  The analysis of the spatial distributions of stars in NGC 5458, 5453, and 5451 showed no statistically significant differences between young and more evolved stars whereas NGC 5461, 5462, and 5449 showed differences that were statistically significant indicating the presence of a spatially varying mean stellar age.

\end{enumerate}

Future work on the subject of massive star population in M101 will address its formation history and the properties of its most luminous stars. Paper II will be a comparison and inter-comparison of the star formation histories in arm, inter-arm, and regions of intense star formation.  Papers III and IV will focus on the properties of the most luminous and variables stars.  We are using the LBT multi-color four year imaging survey of nearby galaxies \citep{Kochanek:2008} to identify the luminous stars in M101 that show evidence for instability from their short-term variability.  Their light curves and photometric properties will be discussed in Paper III.  In the fourth and final paper in this series, we will report on spectroscopy of the luminous variables and other high luminosity stars selected from our catalog.

\acknowledgments Our research on massive stars is supported by the National Science Foundation AST-1019394 (R. Humphreys, P.I.).  All of the data presented in this paper were obtained from the Mikulski Archive for Space Telescopes (MAST). STScI is operated by the Association of Universities for Research in Astronomy, Inc., under NASA contract NAS5-26555. Support for MAST for non-HST data is provided by the NASA Office of Space Science via grant NNX09AF08G and by other grants and contracts.

Facilities: \facility{HST/ACS}

\bibliography{ms_ApJ.bbl}

\end{document}